\documentclass[showpacs,prb]{revtex4}
\usepackage{amsfonts}
\usepackage{graphicx}

\input{tcilatex}

\begin{document}

\title{Mutual-Chern-Simons effective theory of doped antiferromagnets }
\author{ Su-Peng Kou$^{1},$ Xiao-Liang Qi$^{2}$, and Zheng-Yu Weng$^{2}$}
\affiliation{$^{1}$Department of Physics, Beijing Normal University, Beijing, 100875,
China\\
$^{2}$Center for Advanced Study, Tsinghua University, Beijing, 100084, China}

\begin{abstract}
A mutual-Chern-Simons Lagrangian is derived as a minimal field theory
description of the phase-string model for doped antiferromagnets. Such an
effective Lagrangian is shown to retain the full symmetries of parity,
time-reversal, and global SU(2) spin rotation, in contrast to conventional
Chern-Simons theories where first two symmetries are usually broken. Two
ordered phases, i.e.,\emph{\ }antiferromagnetic and superconducting states,
are found at low temperatures as characterized by \textquotedblleft
dual\textquotedblright\ Meissner effects and dual flux quantization
conditions due to the mutual-Chern-Simons gauge structure. A
\textquotedblleft dual\textquotedblright\ confinement in charge/spin degrees
of freedom occurs such that no true spin-charge separation is present in
these ordered phases, but the spin-charge separation/deconfinement serves as
a driving force in the unconventional phase transitions of these ordered
states to disordered states.
\end{abstract}

\maketitle

\section{\protect\bigskip Introduction}

Gauge theory description has become essential in studying doped Mott
insulators. The physical necessity may be traced to the Hilbert space
restriction in a doped Mott insulator. For instance, the high-$T_{c}$
cuprate superconductors at half-filling are believed to be an
antiferromagnetic (AF) Mott insulator \cite{anderson}, in which the charge
sector at low energy is totally frozen up by the Coulomb interaction. After
doping, the low-energy charge degrees of freedom do emerge, but remain
highly restricted in the Hilbert space \cite{anderson}. To characterize such
a Hilbert space restriction, a spin-charge separation description, namely,
by introducing \cite{anderson1,krs,an1} spinless \textquotedblleft
holon\textquotedblright\ of charge $+e$ and neutral spin-$1/2$
\textquotedblleft spinon\textquotedblright\ as the essential \emph{building
blocks} of the restricted Hilbert space, has become an effective and useful
way. Here \textquotedblleft holons\textquotedblright\ and \textquotedblleft
spinons\textquotedblright\ do not necessarily turn out to be true low-lying
elementary excitations in the end, because generally local gauge field(s)
will emerge \cite{fra1,u1} to mediate interactions between these
\textquotedblleft holons\textquotedblright\ and \textquotedblleft
spinons\textquotedblright ,\ and may even lead to the \emph{confinement }of
them if either a true spin-charge separation does not exist or the
decomposition is not done in a correct way. In general, one always ends up
with a gauge theory description for doped Mott insulators where the gauge
interaction can greatly influence the low-energy dynamics of the charge and
spin degrees of freedom.

Several kinds of (2+1)-dimensional gauge theories have been proposed for
doped two-dimensional (2D)\ spin-$1/2$ antiferromagnets related to the high-$%
T_{c}$ cuprates. A $\mathrm{U(1)}$ gauge theory \cite{u1,lee} based on the
slave-boson approach to the $t-J$ model is one of the most intensively
studied. Its gauge structure may be directly visualized by noting the gauge
invariance of the electron operator in the slave-boson decomposition \cite%
{anderson1} 
\begin{equation}
c_{i\sigma }=b_{i}^{\dagger }f_{i\sigma }  \label{slave-boson}
\end{equation}%
under a \textrm{U(1)} transformation: $b_{i}\rightarrow e^{i\theta
_{i}}b_{i} $ and $f_{i\sigma }\rightarrow e^{i\theta _{i}}f_{i\sigma },$
where $b_{i}$ denotes the bosonic \textquotedblleft holon\textquotedblright\
operator and $f_{i\sigma }$ the fermionic \textquotedblleft
spinon\textquotedblright\ operator$.$ Along the same line, the \textrm{SU(2) 
}non-Abelian\textrm{\ }gauge theories\textrm{\ }\cite{su2}\textrm{\ }and%
\textrm{\ Z}$_{2}$ gauge theories \cite{z2} have also been proposed and
studied.\textbf{\ }

The slave-boson approach is considered to be convenient in dealing with the
superconducting (SC) regime but has less advantage in describing the AF
state near half-filling. On the other hand, gauge theories \cite%
{wie,shankar,lee2,weng1,su,ng} based on the slave-fermion, Schwinger boson
decomposition are believed to be useful in studying a lightly doped AF
state. Here the electron operator is written as \cite{aa} 
\begin{equation}
c_{i\sigma }=f_{i}^{\dagger }b_{i\sigma }  \label{slave-f}
\end{equation}%
where $f_{i}$\ denotes the fermionic \textquotedblleft
holon\textquotedblright\ operator and $b_{i\sigma }$\ the bosonic
\textquotedblleft spinon\textquotedblright\ operator. Besides the
slave-boson and slave-fermion decompositions, slave-anyon decompositions
have also been investigated \cite{wiegmann,rod,tik,weng-semion}. Different
gauge structures mentioned above originate from different decompositions
and/or different mean-field decouplings. But a common feature for these
gauge theories is that both \textquotedblleft holon\textquotedblright\ and
\textquotedblleft spinon\textquotedblright\ share the same gauge field.

Recently, a different gauge theory description has been constructed \cite%
{phase-string1} based on a distinctive decomposition of the electron
operator \cite{ps00,phase-string0}

\begin{equation}
c_{i\sigma }=h_{i}^{\dagger }b_{i\sigma }e^{i\hat{\Theta}_{i\sigma }}
\label{ps}
\end{equation}%
which is known as the \emph{bosonization }\cite{ps00} or \emph{phase string
decomposition }\cite{phase-string0} because holon and spinon operators, $%
h_{i}^{\dagger }$ and $b_{i\sigma }$, are both bosonic, with the fermionic
commutations relations of the electron operator being restored by the phase
string operator, $e^{i\hat{\Theta}_{i\sigma }}=(-\sigma )^{i}e^{i\frac{1}{2}%
\left[ \Phi _{i}^{b}-\sigma \Phi _{i}^{h}\right] }$. Here \emph{internal}
gauge invariance appears as \textrm{U(1)}$\times \mathrm{U(1)}$: \ $%
h_{i}\rightarrow e^{i\phi _{i}}h_{i}$ and $\Phi _{i}^{b}\rightarrow \Phi
_{i}^{b}+2\phi _{i};$ $b_{i\sigma }\rightarrow e^{i\sigma \chi
_{i}}b_{i\sigma }$ and $\Phi _{i}^{h}\rightarrow \Phi _{i}^{h}+\chi _{i}$.
Consequently there exist a pair of \textrm{U(1)}$\times \mathrm{U(1)}$ gauge
fields coupling to the holon and spinon fields, respectively, in the
resulting gauge theory, called the phase string model, derived \cite%
{phase-string1} based on the decomposition (\ref{ps}) and the bosonic
resonating-valence-bond (RVB) mean-field saddle-point, where the normal 
\textrm{U(1)} gauge freedom \cite{wang} (like the one in the slave-boson
case) is broken by the mean-field decoupling \cite{remark1}.

In the slave-boson (or slave-fermion) U(1) gauge theory, the external U(1)
gauge field (i.e., the electromagnetic field) couples to \emph{both} holons
and spinons \cite{u1,lee}, thanks to the same \emph{internal} U(1) gauge
field they share. So both holon and spinon carry some fractions of the
electron charge \cite{lee,nayak2}. In contrast, in the phase string model,
the external electromagnetic field only couples to the holon degrees of
freedom, without being directly transferred to the spinon part as the latter
sees a different gauge field. In this sense, the holon carries the full
charge of $+e$\ in the phase string model.

Without a bare kinetic energy, the single \textrm{U(1)} gauge field in the
slave-boson (or slave-fermion) theory fluctuations strongly \cite{lee,weng1}%
, which makes the theory a strong-coupling one. On the other hand, the 
\textrm{U(1)}$\times \mathrm{U(1)}$ gauge fields are topological ones with
their strengths constrained to the densities of two matter fields (see Sec.
II) such that their fluctuations are much more mildly, suitable for a
perturbative treatment. In particular, the no-double-occupancy constraint of
the doped Mott insulator, which is enforced by the violent gauge
fluctuations in the slave-boson (or slave-fermion) theory, is realized in
the phase string model in a quite different way. Namely, the \textrm{U(1)}$%
\times \mathrm{U(1)}$ topological gauge fields will introduce mutual
repulsions between holons and spinons, where holons perceive spinons as
vortices and vice versa. As it is well known, a particle cannot go to the
core of a vortex of its own field where the density of such a matter field
vanishes. In the phase string model such a vortex core of one species is
always occupied by a different species such that the no-double-occupancy is
naturally enforced.

Furthermore, the weak (logarithmic) confinement of spinons and holons at low
energies and low temperatures has been also found \cite{weng-LG,kou,kou1} in
the phase string model, as opposed to the strong confinement in usual 2D
compact \textrm{U(1)} gauge models in slave-boson or slave-fermion theory 
\cite{confine,nayak1}. In the latter, an effective gauge theory may have a
serious infrared divergence \cite{u1,lee} which makes the gauge theory very
difficult to deal with mathematically. The former is usually much more
manageable than the latter in this regard.

However, the Hamiltonian formalism \cite{phase-string1} of the phase string
model, in which a gauge field seen by one species is constrained to the
density (number) of different species, is not very convenient for studies
beyond the mean-field level. In this paper, we shall develop a Lagrangian
(path-integral) formalism of the phase string model. We show that the
effective low-energy Lagrangian describes two matter fields, holon and
spinon, minimally couple to two \emph{different}\textbf{\ }\textrm{U(1)}
gauge fields. These gauge fields have no their own kinetic terms either, but
there is a \emph{mutual-Chern-Simons} term which entangles two gauge fields
together. We call this as a mutual-Chern-Simons description, which
constitutes a minimal field-theory description for the phase string theory.

The gauge structure of such a (2+1)-dimensional mutual-Chern-Simons theory
is very unique in many aspects as compared to the gauge theories proposed
before. We demonstrate that the physical symmetries, which include parity,
time-reversal, and spin rotational symmetries, are precisely preserved in
such an effective theory. By contrast, in usual Chern-Simons (anyon)
theories \cite{laughlin,wwz,wilczek}, the parity and time-reversal
symmetries are explicitly broken, including the mutual-Chern-Simons theory
previously proposed \cite{frac} for describing the double-layer quantum Hall
effect system.

We further show that there exist two low-temperature phases in such a theory
at low doping. One is an AF state which recovers the AF long range order
(AFLRO) of the Heisenberg model at half-filling and may survive at small
doping concentration. The other is an SC state. Two phases are characterized
by dual Meissner effects and dual flux quantization conditions, accompanied
by a dual confinement, which are the direct consequences of the
mutual-Chern-Simons gauge fields interacting with two matter fields when one
of them experiences Bose condensation. Such a mutual duality connecting the
AF and SC states or spin and charge degrees of freedom, is quite different
from the usual duality descriptions proposed \cite{Balents,dhlee,fra,za} for
the cuprate superconductors, where the conventional boson-vortex duality is
used to describe an ordered-disordered transition.

In the SC phase, for example, the Meissner effect and $hc/2e$ flux
quantization are similar to the predictions by a conventional
superconductivity theory, and the spinons are found to be confined such that
to drop out of the physical spectrum. Only \emph{integer} spin excitations,
as composed of confined spinon pairs, are allowed in the bulk state. But as
a unique prediction, a single spinon (an $S=1/2$ moment) does appear in the
center of a magnetic vortex core. It forecasts that the spin
fractionalization will occur in the pseudogap phase, as the latter may be
viewed as the proliferation of the vortex core state above the
superconducting transition $T_{c}$ \cite{weng-LG,ming}.

In the AF phase, on the other hand, the spinon condensation may be viewed as
a two-component \textquotedblleft superfluidity\textquotedblright . The dual
Meissner effect means that a holon is an \textquotedblleft
alien\textquotedblright\ object in the spinon condensate, and the dual flux
quantization condition corresponds to that a meron (vortex) is produced in
the spinon condensate to which a holon must be confined to, just like a
spinon is confined to a magnetic vortex core in the above-mentioned SC
state. As a result, only the \textquotedblleft neutral\textquotedblright\
object of a holon-meron composite, not the holon itself, appears in the
low-energy physical spectrum, which has a dipolar spin configuration at long
distance, coexisting with the AFLRO in a dilute hole concentration regime.

The remainder of the paper is organized as follows. In Sec. II, we briefly
introduce the effective Hamiltonian of the phase string model. In Sec. III,
we first derive the Lagrangian (path-integral) formalism in the lattice
version. Then we obtain the low-energy mutual-Chern-Simons gauge theory
description in the continuum limit. In Sec. IV, we examine the symmetries,
including parity, time-reversal, and spin rotational symmetries, of the
mutual-Chern-Simons theory. In Sec. V, we study two low-temperature ordered
phases based on the mutual-Chern-Simons theory and discuss how holons and
spinons behave in the AF and SC phases, respectively, where dual confinement
of holons/spinons is revealed. Finally, the conclusions are given in Sec. VI.

\section{Phase string theory: A minimal model of doped antiferromagnets}

The phase string theory has been proposed \cite{phase-string0,phase-string1}
as a low-energy effective description of the doped antiferromagnets at low
doping. The \textquotedblleft minimal\textquotedblright\ Hamiltonian of the
phase string theory is composed of two terms, $H_{\mathrm{string}%
}=H_{h}+H_{s},$ in which the charge degrees of freedom are characterized by
the \textquotedblleft holon\textquotedblright\ term

\begin{equation}
H_{h}=-t_{h}\sum_{\langle ij\rangle }\left( e^{iA_{ij}^{s}}\right)
h_{i}^{\dagger }h_{j}+H.c.  \label{hh}
\end{equation}%
where $t_{h}\sim t$ and the \textquotedblleft holon\textquotedblright\
operator, $h_{i}^{\dagger },$ is bosonic; The spin degrees of freedom as
described by the \textquotedblleft spinon\textquotedblright\ term 
\begin{equation}
H_{s}=-J_{s}\sum_{\langle ij\rangle \sigma }\left( e^{i\sigma
A_{ij}^{h}}\right) b_{i\sigma }^{\dagger }b_{j-\sigma }^{\dagger }+H.c.
\label{hs}
\end{equation}%
where $J_{s}\sim J$ and the \textquotedblleft spinon\textquotedblright\
operator, $b_{i\sigma }^{\dagger },$ is also bosonic. Here the gauge fields $%
A_{ij}^{s}$ and $A_{ij}^{h}$ are decided by the topological constraints:

\begin{eqnarray}
\sum_{C}A_{ij}^{s} &=&\pi \sum_{l\in \Sigma _{C}}\left( n_{l\uparrow
}^{b}-n_{l\downarrow }^{b}\right)  \nonumber \\
\sum_{C}A_{ij}^{h} &=&\pi \sum_{l\in \Sigma _{C}}n_{l}^{h}
\label{constraint2}
\end{eqnarray}%
where $n_{l\sigma }^{b}$ and $n_{l}^{h}$ denote the \textquotedblleft
spinon\textquotedblright\ (with index $\sigma $) and \textquotedblleft
holon\textquotedblright\ number operators at site $l$, respectively, and the
path $C$ is an arbitrary loop made of the nearest-neighbor (nn) links with $%
\Sigma _{C}$ denoting the area enclosed by $C.$

Basic features of this model are as follows. At half filling, the gauge
field $A_{ij}^{h}$ can be set to zero in (\ref{hs}) and $H_{s}$ reduces to
the Schwinger-boson mean-field Hamiltonian \cite{aa}, which describes both
the long-range and short-range AF correlations fairly well. Upon doping, $%
A_{ij}^{h}$ is no longer trivial due to constraint (\ref{constraint2}),
which describes that each \textquotedblleft holon\textquotedblright\ behaves
like a $\pi $-fluxoid as felt by the \textquotedblleft
spinons\textquotedblright . Thus, $A_{ij}^{h}$ will play the role of dynamic
frustrations, introduced by doped holes, that acts on the spin degrees of
freedom. Similarly, the \textquotedblleft holons\textquotedblright\ are also
subjected to dynamic frustrations, from the spin background, via the gauge
field $A_{ij}^{s}$ in (\ref{hh}). The spin and charge degrees of freedom are
thus mutually frustrated\ in the phase-string model in terms of two
topological gauge fields, $A_{ij}^{h}$ and $A_{ij}^{s}$.

The phase string model outlined above incorporates, as a minimal model,
three most essential characteristics of the doped antiferromagnets described
by the $t-J$ model. They are: (i) the restricted Hilbert space of doped Mott
insulators, which is characterized by the spin-charge separation formalism
with holons and spinons as basic building blocks; (ii) strong short-range AF
correlations as provided by the bosonic RVB description in (\ref{hs}), which
can naturally grow into an AFLRO state as the doping concentration is
reduced to zero; (iii) the mutual singular influence between the charge and
spin degrees of freedom as represented by two topological gauge fields, $%
A_{ij}^{h}$ and $A_{ij}^{s}$, which mathematically capture the phase string
effect identified \cite{phase-string0} in the $t-J$ model. Such a mutual
interaction has been shown \cite{phase-string1,weng-LG,kou,kou1} to be
responsible for some nontrivial physical properties of the model in close
connection with the high-$T_{c}$ materials.

In the phase string formalism, the spin operators are expressed in terms of
the spinon operators in the following nontrivial form \cite{phase-string0}

\begin{eqnarray}
S_{i}^{z} &=&\frac{1}{2}(b_{i\uparrow }^{\dagger }b_{i\uparrow
}-b_{i\downarrow }^{\dagger }b_{i\downarrow })  \nonumber \\
S_{i}^{+} &=&(S_{i}^{-})^{\dagger }=(-1)^{i}b_{i\uparrow }^{\dagger
}b_{i\downarrow }e^{i\Phi _{i}^{h}}  \label{operators}
\end{eqnarray}%
where the phase $\Phi _{i}^{h}$ appearing in $S_{i}^{\pm }$ can be decided
by the relation $\Phi _{i}^{h}-\Phi _{j}^{h}=2A_{ij}^{h}$ for two nn sites, $%
i$ and $j$, which are not occupied by the holes. Under this definition, the
spin operators as well as the effective Hamiltonian are invariant under the
gauge transformation $b_{i\sigma }\rightarrow b_{i\sigma }e^{i\sigma \phi
_{i}},$ $\Phi _{i}^{h}\rightarrow \Phi _{i}^{h}+2\phi _{i}$, $%
A_{ij}^{h}\rightarrow A_{ij}^{h}+\phi _{i}-\phi _{j}$. The holon-dependent
phase factor $\Phi _{i}^{h}$ in (\ref{operators}) further illustrates the
intrinsic mutual entanglement between spin and charge degrees of freedom in
the phase string theory.

\section{Mutual-Chern-Simons Gauge-Theory Description}

\subsection{Lagrangian Formulation}

The treatment of the Hamiltonian formalism of the phase string model may not
be convenient beyond the mean-field approximation because the gauge fields, $%
A_{ij}^{s}$ and $A_{ij}^{h}$, defined in (\ref{constraint2}), are themselves
operators depending on the dynamics of the matter fields. In order to deal
with the phase string model [(\ref{hh}) and (\ref{hs})] more conveniently, a
Lagrangian (path-integral) formalism will be introduced in this section.

\begin{figure}[tbp]
\begin{center}
\includegraphics[width=2.5in] {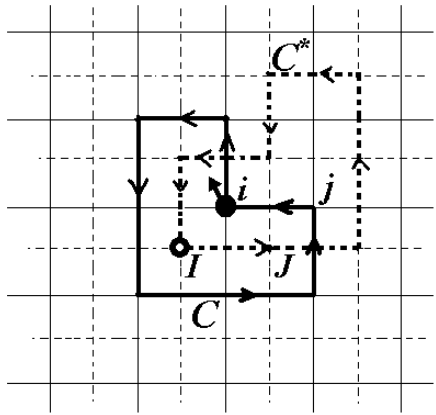}
\end{center}
\caption{A regularization of the contraints in (\protect\ref{constraint2})
by introducing dual lattices is shown. A spinon (denoted by an arrow) and a
holon (denoted by an open circle) stay in dual lattices (solid and dashed
ones, respectively), with the gauge fields $A_{ij}^{h}$ and $A_{IJ}^{s}$
defined on the links of two dual lattices, respectively. The closed loop, $C$
($C^{\ast }$)$,$ of a spinon (holon) can be arbitary without crossing holons
(spinons). See text for the detail.}
\label{duallattice}
\end{figure}

First of all, let us re-express the topological constraint (\ref{constraint2}%
) locally. As pointed out before, the original no-double-occupancy
constraint in the $t-J$ model can be realized in the phase string model by
the mutual repulsion between spinons and holons via $A_{ij}^{h}$ and $%
A_{ij}^{s}$. As a consequence, the closed path $C$ of a holon/spinon in (\ref%
{constraint2}) will not cross spinons/holons and thus effectively avoid a
singularity occurring when a spinon and a holon simultaneously stay at the
same site (as each spinon/holon carries a $\pi $-fluxoid seen by a
holon/spinon). Following this, then, it is physically reasonable to
implement a regularization in the topological constraint (\ref{constraint2})
by introducing two sets of dual square lattices, respectively, for spinons
and holons to stay, as illustrated in Fig. \ref{duallattice}. In this way, a
closed path $C$ for spinon and a $C^{\ast }$ for holon on different lattices
can be arbitrary without worrying to cross the opposite species, either
holons or spinons. Presumably no important low-energy physics will get lost
by such a \emph{local} regularization. 

Here and below, the minuscule (majuscule) Latin letters $i,j$ ($I,J$) will
be used to label the dual lattice sites for spinons (holons). The Greek
letters $\alpha $, $\beta $, $\gamma $ will be used for 2D spatial indices, $%
1$ and $2$, while $\mu $, $\nu $, $\lambda $ for the three dimensional
space-time indices, $0,1,2$. Then the topological constraint (\ref%
{constraint2}) can be re-expressed in a compact form as follows: 
\begin{eqnarray}
\epsilon ^{\alpha \beta }\Delta _{\alpha }A_{\beta }^{h}(i) &=&\pi n_{I}^{h}
\nonumber \\
\epsilon ^{\alpha \beta }\Delta _{\alpha }A_{\beta }^{s}(I) &=&\pi
\sum_{\sigma }\sigma n_{i\sigma }^{b},  \label{constraint}
\end{eqnarray}%
in which the link fields $A_{\alpha }^{h}(i)\equiv A_{i+\hat{\alpha},i}^{h}$
and $A_{\alpha }^{s}(I)\equiv A_{I,I-\hat{\alpha}}^{s}$, with $\alpha =x$, $%
y,$ and the difference operators, $\Delta _{\alpha },$ on the two sets of
the dual lattices are defined by $\Delta _{\alpha }f(i)=f(i+\hat{\alpha}%
)-f(i)$ and $\Delta _{\alpha }f(I)=f(I)-f(I-\hat{\alpha}),$ respectively.
Note the slightly different definitions of link variables and lattice
difference operators on two dual lattices, so as to keep the symmetric forms
in (\ref{constraint}).

In the path-integral formulation, the topological constraint (\ref%
{constraint}) can be enforced by introducing two Lagrangian multipliers, $%
A_{0}^{h}(i)$ and $A_{0}^{s}(I),$ as follows 
\begin{equation}
L_{\mathrm{constr}}=-i\sum_{I}A_{0}^{s}(I)\left[ n_{I}^{h}-\frac{1}{\pi }%
\epsilon ^{\alpha \beta }\Delta _{\alpha }A_{\beta }^{h}(i)\right]
-i\sum_{i}A_{0}^{h}(i)\left[ \sum_{\sigma }\sigma n_{i\sigma }^{b}-\frac{1}{%
\pi }\epsilon ^{\alpha \beta }\Delta _{\alpha }A_{\beta }^{s}(I)\right] .
\label{Lconstr}
\end{equation}%
Once the topological constraint is implemented by the Lagrangian
multipliers, the gauge fields, $A_{I\alpha }^{s}$ and $A_{i\alpha }^{h},$
can be treated as \emph{independent} gauge variables in the Lagrangian
formalism. In order to get the correct form of the Lagrangian for this
system, we need to first identify the canonical momenta of the gauge fields, 
$A_{ij}^{h}$ and $A_{IJ}^{s}.$

It is helpful to consider the continuity equation for the holon density: 
\begin{equation}
\partial _{t}n_{I}^{h}+\Delta ^{\alpha }J_{I+\hat{\alpha},I}^{h}=0.
\label{contin}
\end{equation}%
Using the topological constraint in (\ref{constraint}) and the definition of
the conserved holon current, $J_{I+\hat{\alpha},I}^{h}=-\frac{\delta H_{%
\mathrm{string}}}{\delta A_{I+\hat{\alpha},I }^{s}},$ one gets 
\begin{equation}
\partial _{t}\left[ \frac{1}{\pi }\epsilon ^{\alpha \beta }\Delta _{\alpha
}A_{\beta }^{h}(i)\right] +\Delta ^{\alpha }\left[ -\frac{\delta H_{\mathrm{%
string}}}{\delta A_{I+\hat{\alpha},I}^{s}}\right] =0  \label{Motion}
\end{equation}%
such that [under a proper gauge choice of $A_{\beta }^{h}(i)$] 
\begin{equation}
\partial _{t}A_{\beta }^{h}(i)=\frac{\delta H_{\mathrm{string}}}{\delta
\left( -\pi ^{-1}\epsilon ^{\beta \gamma }A_{\gamma }^{s}(I)\right) }.
\label{Motion2}
\end{equation}

Equation (\ref{Motion2}) is just the canonical equation of motion for $%
A_{\beta }^{h}(i)$, and one can thus identify the canonical momentum $\Pi
_{\beta }^{h}(i)=-\frac{1}{\pi }\epsilon ^{\beta \gamma }A_{\gamma }^{s}(I)$%
. In other words, the spatial components of the gauge fields $A^{h}$ and $%
A^{s}$ are canonically conjugate to each other. (The temporal components, $%
A_{0}^{s}$ and $A_{0}^{h},$ have no canonical momenta since they do not have
independent dynamics in the above formulation).

Following the standard canonical quantization procedure, the \emph{Euclidean}
Lagrangian (with the Wick rotation $t\rightarrow -i\tau $) of this system
can be derived straightforwardly as follows%
\begin{eqnarray*}
L_{\mathrm{string}} &=&\sum_{i}\Pi _{\alpha }^{h}(i)(-i)\partial
_{0}A_{\alpha }^{h}(i)+\sum_{i}b_{i\sigma }^{\dagger }\partial
_{0}b_{i\sigma }+\sum_{I}h_{I}^{\dagger }\partial _{0}h_{I}+H_{\mathrm{string%
}}+L_{\mathrm{constr}} \\
&\equiv &L_{h}+L_{s}+L_{CS}
\end{eqnarray*}%
where 
\begin{eqnarray}
L_{h} &=&\sum_{I}h_{I}^{\dagger }\left[ \partial _{0}-iA_{0}^{s}(I)\right]
h_{I}-t_{h}\sum_{\left\langle IJ\right\rangle }\left( h_{I}^{\dagger
}e^{iA_{IJ}^{s}}h_{J}+h.c.\right) +\mu \left( \sum_{I}h_{I}^{\dagger
}h_{I}-N\delta \right)  \nonumber \\
L_{s} &=&\sum_{i\sigma }b_{i\sigma }^{\dagger }\left[ \partial _{0}-i\sigma
A_{0}^{h}(i)\right] b_{i\sigma }-J_{s}\sum_{\left\langle ij\right\rangle
\sigma }\left( b_{i\sigma }^{\dagger }e^{i\sigma A_{ij}^{h}}b_{j-\sigma
}^{\dagger }+h.c.\right) +\lambda \left( \sum_{i\sigma }b_{i\sigma
}^{\dagger }b_{i\sigma }-N\left( 1-\delta \right) \right)  \nonumber \\
L_{CS} &=&\frac{i}{\pi }\sum_{I}\epsilon ^{\mu \nu \lambda }A_{\mu
}^{s}(I)\partial _{\nu }A_{\lambda }^{h}(i)  \label{Lagrangian-CS}
\end{eqnarray}%
with $\partial _{0}\equiv \partial _{\tau }$. Note that one can also use a
procedure similar to (\ref{contin})-(\ref{Motion2}) to define a conjugate
field $\Pi _{\alpha }^{s}(I)=-\frac{1}{\pi }\epsilon ^{\alpha \beta
}A_{\beta }^{h}(i)$ for $A_{\alpha }^{s}(I)$ and the resulting Lagrangian
remains the same as above.

Therefore, the Lagrangian formalism of the phase string model describes that
the two matter fields, bosonic spinons and holons, are minimally coupled to $%
U(1)\times U(1)$ gauge fields, $A_{\mu }^{s}$ and $A_{\mu }^{h}$, whose
gauge structure is decided by the mutual-Chern-Simons term $L_{CS}$ in (\ref%
{Lagrangian-CS}). In the following, we shall further derive the
long-wavelength, low-energy effective Lagrangian based on such a lattice
model.

\subsection{Low-energy effective theory}

The Lagrangian (\ref{Lagrangian-CS}) is written in a lattice form. It can be
further simplified and reduced to a continuum version in the
long-wavelength, low-energy limit. The procedure given below is quite
standard and straightforward.

Let us first consider the spinon Lagrangian $L_{s}$, in which some careful
treatment is needed in taking the continuum limit$.$ We shall derive its
low-energy action in the $CP(1)$ formalism \cite{read} by integrating out
the short-range \emph{ferromagnetic} fluctuations.

First of all, we divide the square lattice into two sublattices, $A$ and $B$%
, and redefine the spinon operator $b_{i\sigma }$ at $B$ sublattice as $\bar{%
b}_{i\sigma }$. Then $L_{s}$ in (\ref{Lagrangian-CS}) can be rewritten as 
\begin{eqnarray}
L_{s} &=&\sum_{i\in A,\sigma }b_{i\sigma }^{\dagger }(\partial _{0}-i\sigma
A_{0}^{h})b_{i\sigma }+\sum_{i\in B,\sigma }\bar{b}_{i\sigma }^{\dagger
}(\partial _{0}-i\sigma A_{0}^{h})\bar{b}_{i\sigma }  \nonumber \\
&&-J_{s}\sum_{i\in A,j=\text{nn(}i),\sigma }\left( b_{i\sigma }^{\dagger
}e^{i\sigma A_{ij}^{h}}\bar{b}_{j-\sigma }^{\dagger }+h.c.\right)  \nonumber
\\
&&+\lambda \left( \sum_{i\in A,\sigma }b_{i\sigma }^{\dagger }b_{i\sigma
}+\sum_{i\in B,\sigma }\bar{b}_{i\sigma }^{\dagger }\bar{b}_{i\sigma
}-N\left( 1-\delta \right) \right) .  \label{LB}
\end{eqnarray}

As usual, we introduce the following continuum fields 
\begin{eqnarray}
b_{i\sigma } &=&z_{\sigma }(\mathbf{r}_{i})+\pi _{\sigma }(\mathbf{r}_{i}) 
\nonumber \\
\bar{b}_{i+\hat{\eta},-\sigma }^{\dagger } &=&z_{\sigma }(\mathbf{r}_{i}+%
\hat{\eta}a)-\pi _{\sigma }(\mathbf{r}_{i}+\hat{\eta}a)  \label{CP1}
\end{eqnarray}%
in which $i\in A$, $\hat{\eta}=\hat{x},\hat{y}$, and $a$ is the lattice
constant. Then, by expressing the Lagrangian (\ref{LB}) in terms of $%
z_{\sigma }$ and $\pi _{\sigma }$ and taking the continuum limit $%
a\rightarrow 0$ with $A_{\alpha }^{h}(i)\rightarrow aA_{\alpha }^{h}(\mathbf{%
r})$, we obtain $L_{s}=\int d^{2}\mathbf{r}\mathcal{L}_{s}$, in which 
\begin{eqnarray}
\mathcal{L}_{s} &=&\sum_{\sigma }\left[ J_{s}\left\vert \left( \partial
_{\alpha }-i\sigma A_{\alpha }^{h}\right) z_{\sigma }\right\vert
^{2}+a^{-2}\left( \lambda -4J_{s}\right) \left\vert z_{\sigma }\right\vert
^{2}\right] -\lambda a^{-2}\left( 1-\delta \right)  \nonumber \\
&&+\sum_{\sigma }\left[ -J_{s}\left\vert \left( \partial _{\alpha }-i\sigma
A_{\alpha }^{h}\right) \pi _{\sigma }\right\vert ^{2}+a^{-2}\left( \lambda
+4J_{s}\right) \left\vert \pi _{\sigma }\right\vert ^{2}\right]  \nonumber \\
&&+a^{-2}\sum_{\sigma }\left[ \pi _{\sigma }^{\ast }\left( \partial
_{0}-i\sigma A_{0}^{h}\right) z_{\sigma }-\pi _{\sigma }\left( \partial
_{0}+i\sigma A_{0}^{h}\right) z_{\sigma }^{\ast }\right]
\end{eqnarray}%
By further integrating out the high-energy field $\pi ^{\sigma },$ we arrive
at 
\begin{equation}
\mathcal{L}_{s}=\sum_{\sigma }\left( \frac{a^{-2}}{\lambda +4J_{s}}%
\left\vert \left( \partial _{0}-i\sigma A_{0}^{h}\right) z_{\sigma
}\right\vert ^{2}+J_{s}\left\vert \left( \partial _{\alpha }-i\sigma
A_{\alpha }^{h}\right) z_{\sigma }\right\vert ^{2}+a^{-2}\left( \lambda
-4J_{s}\right) \left\vert z_{\sigma }\right\vert ^{2}\right) -\lambda
a^{-2}\left( 1-\delta \right)
\end{equation}

Define the spin-wave velocity $c_{s}=\sqrt{J_{s}(\lambda +4J_{s})}a$ and
redefine the temporal components: $\tau \rightarrow c_{s}x_{0},$ $%
A_{0}^{h}\rightarrow c_{s}A_{0}^{h}$, the low-energy effective action for
the spinons can be finally written as 
\begin{equation}
S_{s}=\int d^{2}\mathbf{r}\int_{0}^{c_{s}\beta }dx_{0}\frac{1}{2g}\left[
|(\partial _{\mu }-i\sigma A_{\mu }^{h})z_{\sigma }|^{2}+m_{s}^{2}|z_{\sigma
}|^{2}\right] ,  \label{eff}
\end{equation}%
Here the summations over $\mu =0,1,2$ and $\sigma =\uparrow ,\downarrow $
are omitted for simplicity and the constant term $\lambda a^{-2}\left(
1-\delta \right) $ is also dropped. The coupling constant $%
g=c_{s}/2J_{s}\left( 1-\delta \right) ,$ and the mass $m_{s}=c_{s}^{-1}\sqrt{%
\lambda ^{2}-16J_{s}^{2}}$, in which $\lambda $ is decided by the spinon
number constraint $\int d^{2}x\sum_{\sigma }\left\vert z_{\sigma
}\right\vert ^{2}=Na^{2}$. Note that here the $z_{\sigma }$ field has been
rescaled in the last step such that $\sum_{\sigma }\left\vert z_{\sigma
}\right\vert ^{2}$ keeps to be $1$ per site on average even at finite
doping. Therefore, in its final form, the long-wavelength theory for spinons
consists of a massive, spin-1/2, and relativistic bosonic $z_{\sigma }$
(spinon) coupled to a $U(1)$ gauge field $A_{\mu }^{h}$.

The continuum versions of $L_{h}$ and $L_{CS}$ can be more straightforwardly
obtained by directly taking the continuum limit $a\rightarrow 0$, with $%
A_{\alpha }^{s}(I)\rightarrow aA_{\alpha }^{s}(\mathbf{r})$, $%
A_{0}^{s}(I)\rightarrow c_{s}A_{0}^{s}(\mathbf{r})$, and $h_{I}\rightarrow
ah(\mathbf{r})$. The final form of the partition function can be written in
the compact form

\[
Z=\int DhDz_{\uparrow }Dz_{\downarrow }DA^{s}DA^{h}\exp \left(
-\int_{0}^{c_{s}\beta }dx_{0}\int d^{2}\mathbf{r}\mathcal{L}_{\mathrm{eff}%
}\right) 
\]%
in which 
\begin{equation}
\mathcal{L}_{\mathrm{eff}}=\mathcal{L}_{h}+\mathcal{L}_{s}+\mathcal{L}_{CS}
\label{effectiveL}
\end{equation}%
with%
\begin{eqnarray}
\mathcal{L}_{h} &=&h^{\dagger }\left[ \partial _{0}-i(A_{0}^{s}+eA_{0}^{e})%
\right] h+h^{\dagger }\frac{\left( -i\partial _{\alpha }-A_{\alpha
}^{s}-eA_{\alpha }^{e}\right) ^{2}}{2m_{h}}h  \nonumber \\
\mathcal{L}_{s} &=&\frac{1}{2g}\left[ |(\partial _{\mu }-i\sigma A_{\mu
}^{h})z_{\sigma }|^{2}+m_{s}^{2}|z_{\sigma }|^{2}\right]  \nonumber \\
\mathcal{L}_{CS} &=&\frac{i}{\pi }\epsilon ^{\mu \nu \lambda }A_{\mu
}^{s}\partial _{\nu }A_{\lambda }^{h}  \label{continous}
\end{eqnarray}%
where in the holon Lagrangian density $\mathcal{L}_{h}$, $m_{h}$ $\simeq
(2t_{h}a^{2})^{-1},$ $A_{\mu }^{e}$ is the vector potential of the external
electromagnetic field, and $-e$ is the electron electric charge. Note that
the chemical potential $\mu $ in $\mathcal{L}_{h}$ has been absorbed into $%
iA_{0}^{s}$ for simplicity.

The Lagrangians in (\ref{continous}) constitute our final low-energy
effective theory. They describe two matter fields, holons and spinons,
minimally coupled to a pair of \textrm{U(1)}$\times \mathrm{U(1)}$ gauge
fields, $A_{\mu }^{s}$ and $A_{\mu }^{h}$. The latter do not have their own
kinetic energies, but are mutually \textquotedblleft
entangled\textquotedblright\ by the mutual-Chern-Simons term $\mathcal{L}%
_{CS}$. Such a mutual-Chern-Simons term has been previously proposed \cite%
{frac} for describing the double-layer quantum Hall effect system. But here
due to the fact that $A_{\mu }^{h}$ couples to up/down spins with opposite
\textquotedblleft charges\textquotedblright\ in $\mathcal{L}_{s}$, the
parity and time-reversal symmetries are explicitly retained (see below). The
external electromagnetic field, $A_{\mu }^{e}$, only directly couples to the
holon field, indicating that the latter is the primary charge carrier
(consistent with the definition of the holon). This is in contrast to the
usual \textrm{U(1)} gauge theory based on the slave-boson approach, in which
both holon and spinon share the external electromagnetic field as if each of
them carriers a fractional part of the charge $e$ (as the result that both
of them see the same internal \textrm{U(1)} gauge field).

In the following section, we shall carefully examine the symmetries of this
effective Lagrangian with a particular attention to the parity,
time-reversal, and spin \textrm{SU(2) }rotational symmetries.

\section{Symmetries}

The symmetries of the present mutual-Chern-Simons Lagrangian will be studied
in this section. The following discussions will be based on the low-energy
effective Lagrangian (\ref{continous}), although all of them can be easily
generalized to the lattice formalism in (\ref{Lagrangian-CS}).

First of all, we note that the $\mathrm{U(1)_{charge}}\times \mathrm{%
U(1)_{S_{z}}}$ gauge invariance of $S_{\mathrm{eff}}$ is obvious according
to (\ref{continous}). Consequently, the global $\mathrm{U(1)_{charge}}$
invariance of the holons ensures the conservation of the electromagnetic
charge in this system. Also straightforward is the translational invariance
in (2+1)-dimensions. In the following, we shall mainly focus on the parity,
time-reversal, and spin rotational symmetries, and show that they are
explicitly retained in the present mutual-Chern-Simons gauge theory, in
contrast to ordinary Chern-Simons theories in which the parity and
time-reversal symmetries are usually broken.

\subsection{Parity}

In (2+1)-dimensions, the parity transformation is defined as a reflection
with regard to a spatial axis, \emph{e.g.}, 
\begin{equation}
x\rightarrow -x,\text{ }y\rightarrow y,\text{ }\tau \rightarrow \tau
\label{P}
\end{equation}

It is straightforward to verify that the effective Lagrangians, $\mathcal{L}%
_{h}$, $\mathcal{L}_{s},$ and $\mathcal{L}_{CS},$ remain invariant,
respectively, under the parity transformation (\ref{P}), if the matter
fields and gauge fields transform under (\ref{P}) as follows

\begin{eqnarray}
z_{\sigma } &\rightarrow &z_{-\sigma },\text{ \ }h\rightarrow h  \nonumber \\
A_{0}^{h} &\rightarrow &-A_{0}^{h},\text{ }A_{x}^{h}\rightarrow A_{x}^{h},%
\text{ }A_{y}^{h}\rightarrow -A_{y}^{h}  \nonumber \\
A_{0}^{s} &\rightarrow &A_{0}^{s},\text{ }A_{x}^{s}\rightarrow -A_{x}^{s},%
\text{ }A_{y}^{s}\rightarrow A_{y}^{s}  \label{parity}
\end{eqnarray}

The parity transformations of the fields in (\ref{parity}) can be determined
as follows. For example, according to the property of angular momenta, a
spin should transform as an axial vector, namely, $S_{x}\rightarrow S_{x},$ $%
S_{y}\rightarrow -S_{y},S_{z}\rightarrow -S_{z}$ under the parity
transformation (\ref{P}). Thus the transformation of the $CP(1)$ field $%
z_{\sigma }$ should be $z_{\sigma }\rightarrow z_{-\sigma }$. On the other
hand, the gauge field $A^{h}$ transforms as an axial vector and $A^{s}$ as a
polar vector in (\ref{parity}). Indeed, in order to keep the invariance of $%
\mathcal{L}_{h}$ and $\mathcal{L}_{s}$, the parity of $A_{\mu }^{s}$ and $%
A_{\mu }^{h}$ should be identical to the charge current $j_{\mu
}^{h}=-\delta \mathcal{L}_{h}/\delta A_{\mu }^{s}$ and spin current $j_{\mu
}^{s}=-\delta \mathcal{L}_{s}/\delta A_{\mu }^{h}$, respectively.
Furthermore, the parity transformations of $A_{\mu }^{s}$ and $A_{\mu }^{h}$
are also consistent with the classical equations of motion for the
Chern-Simons fields obtained based on (\ref{continous}): 
\begin{equation}
j_{\mu }^{s}=\frac{i}{\pi }\epsilon ^{\mu \nu \lambda }\partial _{\nu
}A_{\lambda }^{s},\text{ }j_{\mu }^{h}=\frac{i}{\pi }\epsilon ^{\mu \nu
\lambda }\partial _{\nu }A_{\lambda }^{h}.
\end{equation}%
The parity invariance of the mutual Chern-Simons term is also related to the
fact that the gauge field $A^{h}$ transforms as an axial vector and $A^{s}$
as a polar vector, in contrast to an ordinary \textrm{U(1)} Chern-Simons
theory.

\subsection{Time-reversal}

Under the time-reversal transformation, 
\begin{equation}
\tau \rightarrow -\tau ,\text{ }r_{\alpha }\rightarrow r_{\alpha },
\label{T}
\end{equation}%
the $z_{\sigma }$ and $h$ field will transform as usual spinor and scalar
fields, respectively. Using the same procedure as given above in the parity
transformation, we can determine 
\begin{eqnarray}
z_{\sigma } &\rightarrow &\sigma z_{-\sigma }^{\ast },\text{ }h\rightarrow
h^{\ast }  \nonumber \\
A_{0}^{h} &\rightarrow &-A_{0}^{h},\text{ }A_{\alpha }^{h}\rightarrow
A_{\alpha }^{h}  \nonumber \\
A_{0}^{s} &\rightarrow &A_{0}^{s},\text{ }A_{\alpha }^{s}\rightarrow
-A_{\alpha }^{s}  \label{timereversal}
\end{eqnarray}%
under the time-reversal transformation (\ref{T}). It can be then easily
checked that the Lagrangian $\mathcal{L}_{\mathrm{eff}}=\mathcal{L}_{h}+%
\mathcal{L}_{s}+\mathcal{L}_{CS}$ is also invariant under the time-reversal
transformation.

The parity and time-reversal invariances of the mutual Chern-Simons
Lagrangian (\ref{continous}) are in sharp contrast to the violations of
both, separately, in an ordinary $\mathrm{U(1)}$ Chern-Simons theory. As
noted above, $A^{h}$ as an axial vector and $A^{s}$ as a polar vector in the
mutual $\mathrm{U(1)\times U(1)}$ Chern-Simons theory are the key for the
restoration of the symmetries. Note that the charge conjugate symmetry is
meaningless here since the holon Lagrangian $\mathcal{L}_{h}$ is
non-relativistic and anti-holons are not well defined.

\subsection{Spin SU(2) rotation}

The demonstration of the global spin $\mathrm{SU(2)}$ symmetry in the
present formulation is less straightforward than the other symmetries
discussed above. The underlying reason is that the spin operators are
expressed in an unconventional way in terms of the $b_{i}\equiv
(b_{i\uparrow },b_{i\downarrow })^{T}$ doublet according to (\ref{operators}%
).

Let us consider a global \textrm{SU(2) }spin rotation defined by $U=\exp
\left( i\mathbf{\theta }\cdot \mathbf{S}\right) .$ In terms of (\ref%
{operators}), one finds $U^{-1}b_{i}U=(\sigma _{3})^{i}e^{-i\sigma _{3}\Phi
_{i}^{h}/2}e^{i\mathbf{\theta }\cdot \mathbf{\sigma }/2}e^{i\sigma _{3}\Phi
_{i}^{h}/2}(\sigma _{3})^{i}b_{i}$. Correspondingly, according to the
definition of the $CP(1)$ fields in (\ref{CP1}), the doublet $z=(z_{\uparrow
},z_{\downarrow })^{T}$ under the $SU(2)$ rotation $\mathrm{U}$ is given by 
\begin{equation}
U^{-1}z(\mathbf{r},\tau )U=e^{i\sigma _{3}\Phi ^{h}(\mathbf{r},\tau )/2}e^{i%
\mathbf{\sigma }\cdot \mathbf{\theta }/2}e^{-i\sigma _{3}\Phi ^{h}(\mathbf{r}%
,\tau )/2}z(\mathbf{r},\tau )  \label{su2}
\end{equation}%
in which 
\begin{equation}
\partial _{\mu }\Phi ^{h}(\mathbf{r},\tau )=2A_{\mu }^{h}(\mathbf{r},\tau ).
\label{su2-1}
\end{equation}%
(Note that in the Hamiltonian formalism, the single-valueness of $\Phi
_{i}^{h}$ in the spin operators (\ref{operators}) is ensured by the
topological constraint on $A_{ij}^{h}$ according to (\ref{constraint2}). In
the path-integral formalism, $\Phi ^{h}$ is determined by (\ref{su2-1}), and
we show in Appendix A that to have a finite contribution to the partition
function, $\Phi ^{h}$ must still satisfy the single-valueness constraint: $%
\Delta \Phi ^{h}|_{C}\equiv \oint_{C}\partial _{\mu }\Phi ^{h}dx_{\mu
}=\oint_{C}2A_{\mu }^{h}dx_{\mu }=2n\pi ,$ with $n\in \mathbb{Z}$ for an
arbitrary loop $C.)$

The spinon Lagrangian $\mathcal{L}_{s}$ can be rewritten as 
\begin{equation}
\mathcal{L}_{s}=\frac{1}{2g}\left\{ \left( D_{\mu }z\right) ^{\dagger
}D_{\mu }z+m_{s}^{2}z^{\dagger }z\right\}  \label{covcons}
\end{equation}%
in which $D_{\mu }z\equiv \left( \partial _{\mu }-i\sigma _{3}A_{\mu
}^{h}\right) z=\partial _{\mu }\left( e^{-i\sigma _{3}\Phi ^{h}/2}z\right) $%
. Under the transformation (\ref{su2}), $D_{\mu }z$ transforms as 
\[
U^{-1}\left( D_{\mu }z\right) U=e^{i\mathbf{\sigma }\cdot \mathbf{\theta }%
/2}D_{\mu }z. 
\]%
Namely, $D_{\mu }z$ transforms as the basic representation of the $\mathrm{%
SU(2)}$ group, and the SU(2) invariance of the Lagrangian (\ref{covcons}) is
proved. Independent of $z(\mathbf{r},\tau )$, Lagrangians $\mathcal{L}_{h}$
and $\mathcal{L}_{CS}$ are obviously invariant. Therefore, the global spin $%
SU(2)$ symmetry is indeed preserved in the present mutual-Chern-Simons
theory.

\section{Two Ordered Phases at Low Temperatures}

\subsection{AF phase at low doping}

\subsubsection{Half-filling}

Let us first consider Lagrangian (\ref{effectiveL}) at half filling. Without
the presence of holons, one can find $A_{\mu }^{h}=0$ and $\mathcal{L}_{%
\mathrm{eff}}$ reduces to a \textrm{CP(1)} model 
\begin{equation}
\mathcal{L}_{\mathrm{eff}}\rightarrow \mathcal{L}_{s}=\frac{1}{2g}\left\{
|\partial _{\mu }z|^{2}+m_{s}^{2}|z|^{2}\right\} .  \label{half-filling}
\end{equation}

The saddle-point solution of (\ref{half-filling}) can be obtained by a
standard procedure after integrating out the \textrm{CP(1)} $z$-field and
then minimizing the resulting action with regard to $m_{s}^{2}$ (here the
constant term $-m_{s}^{2}/2g$ previously dropped in $\mathcal{L}_{s}$ has to
be included) as follows \cite{cha,tri,sach} 
\begin{equation}
gT\sum_{\omega _{n}}\int \frac{d^{2}\mathbf{k}}{4\pi ^{2}}\frac{1}{\mathbf{k}%
^{2}+\omega _{n}^{2}+m_{s}^{2}}=1  \label{ms1}
\end{equation}%
where $\omega _{n}=2\pi nT$, $n=$ integers. With a proper regularization 
\cite{tri} in (\ref{ms1}), the mass gap $m_{s}$ can be determined at small $%
T $ as 
\begin{equation}
m_{s}\approx T\exp (-\frac{2\pi }{T}\frac{1}{\tilde{g}}).  \label{ms}
\end{equation}%
in the so-called renormalized classical region, where $\frac{1}{\tilde{g}}%
\equiv \frac{1}{g}-\frac{1}{g_{c}}>0$ (here $g_{c}=\frac{4\pi }{\Lambda }$
with $\Lambda $ denoting a cutoff parameter in the regularization).

At $T=0$, the mass gap $m_{s}=0,$ and a Bose condensation takes place in the
ground state with $\left\langle z\right\rangle \mathbf{\neq }0$,\textbf{\ }%
corresponding to an AFLRO lying in the x-y plane:\textbf{\ }$\left\langle
S_{i}^{+}\right\rangle =(-1)^{i}<z_{\uparrow }><z_{\downarrow }>,$\textbf{\ }%
which can be easily destroyed by thermal fluctuations at any finite
temperatures as indicated by $m_{s}>0$ according to (\ref{ms}).

The energy scale of the mass gap $m_{s}$ is always much smaller than the
temperature, \emph{i.e., }$m_{s}\ll T,$ at $T\ll \frac{1}{\tilde{g}}.$ Thus, 
$\omega _{n}=2\pi nT$ $(n\geq 1)$ is usually much larger than the mass gap,
which means that the quantum fluctuations will become negligible in a
sufficiently long wavelength and low energy regime, where one may only
consider the purely static (semiclassical) fluctuations. In the region of $%
m_{s}<k<c_{s}\beta $, the effective Lagrangian of the \textrm{CP(1)} field
will lose the Lorentz invariance and becomes 
\begin{equation}
\mathcal{L}_{s}\approx \frac{1}{2\tilde{g}}|\mathbf{\nabla }z|^{2}.
\label{lsclass}
\end{equation}%
Such an effective Lagrangian can be also obtained in the renormalized
classical region by using the \textrm{O(3)} nonlinear $\sigma $ model \cite%
{cha}.

\subsubsection{\protect\bigskip Low doping}

In a sufficiently small concentration of holes, if the AFLRO or the Bose
condensation of the \textrm{CP(1)} spinor fields persists, then the
renormalized classical Lagrangian (\ref{lsclass}) remains applicable, which
should be simply modified to couple to the gauge field $\mathbf{A}^{h}$
according to (\ref{continous}) as follows$:$ 
\begin{equation}
\mathcal{L}_{s}=\frac{1}{2\tilde{g}}|(\mathbf{\nabla }-i\sigma _{3}\mathbf{A}%
^{h})z|^{2}.
\end{equation}%
On the other hand, holons are coupled to $A_{\mu }^{s}$ in $\mathcal{L}_{h}$%
, and two gauge fields are then entangled by the mutual-Chern-Simons term $%
\mathcal{L}_{CS}$ [see\emph{\ }(\ref{continous})], which can be rewritten,
up to a boundary term, as 
\begin{equation}
\mathcal{L}_{CS}=-\frac{i}{\pi }\mathbf{A}^{h}\cdot (\mathbf{E}^{s}\times 
\mathbf{\hat{z}})+\frac{i}{\pi }A_{0}^{h}B^{s}  \label{effAFM}
\end{equation}%
where we introduce $\mathbf{E}^{s}\equiv \partial _{0}\mathbf{A}^{s}-\nabla
A_{0}^{s}$ as the \textquotedblleft electric field\textquotedblright\
strength for $A_{\mu }^{s}$ and $B^{s}=\nabla \times \mathbf{A}^{s}\cdot 
\mathbf{\hat{z}}$ as its \textquotedblleft magnetic field\textquotedblright\
strength. By integrating out $A_{\mu }^{h},$ then, the spin dynamics will
become entangled\ with the holon dynamics as shown below.

First of all, the integration over $A_{0}^{h}$ will simply lead to $B^{s}=0$%
. In the following, one may then choose a proper gauge: $\mathbf{A}^{s}=0$
and $\mathbf{E}^{s}=-\nabla A_{0}^{s}.$ Next, by using%
\[
|(\mathbf{\nabla }-i\sigma _{3}\mathbf{A}^{h})z|^{2}=|\mathbf{\nabla }\tilde{%
z}|^{2}+2\mathbf{A}^{h}\cdot \mathbf{v}^{s}+\left( \mathbf{A}^{h}\right)
^{2}\left\vert \tilde{z}\right\vert ^{2} 
\]%
with $\tilde{z}\equiv \left( z_{\uparrow },z_{\downarrow }^{\ast }\right) ^{%
\mathrm{T}}$ and $\mathbf{v}^{s}\equiv \frac{i}{2}\left( \widetilde{z}%
^{\dagger }\nabla \widetilde{z}-\nabla \widetilde{z}^{\dagger }\widetilde{z}%
\right) ,$ one has

\[
\mathcal{L}_{s}+\mathcal{L}_{CS}=\frac{1}{2\tilde{g}}|\mathbf{\nabla }\tilde{%
z}|^{2}+\mathbf{A}^{h}\cdot \left( \frac{1}{\tilde{g}}\mathbf{v}^{s}-\frac{i%
}{\pi }\mathbf{E}^{s}\times \mathbf{\hat{z}}\right) +\frac{1}{2\tilde{g}}%
\left( \mathbf{A}^{h}\right) ^{2}, 
\]%
under the constraint $\left\vert \tilde{z}\right\vert ^{2}=1,$ which, after
integrating out $\mathbf{A}^{h}$, arrives at%
\begin{eqnarray}
&&\frac{1}{2\tilde{g}}|\mathbf{\nabla }\tilde{z}|^{2}-\frac{\tilde{g}}{2}%
\left( \frac{1}{\tilde{g}}\mathbf{v}^{s}-\frac{i}{\pi }\mathbf{E}^{s}\times 
\mathbf{\hat{z}}\right) ^{2}  \nonumber \\
&&\mathbf{=}\frac{1}{2\tilde{g}}\left( |\mathbf{\nabla }\tilde{z}%
|^{2}-\left\vert \mathbf{v}^{s}\right\vert ^{2}\right) +\frac{\tilde{g}}{%
2\pi ^{2}}(\mathbf{E}^{s})^{2}+\frac{i}{\pi }(\mathbf{E}^{s}\times \mathbf{%
\hat{z}})\cdot \mathbf{v}^{s}\mathbf{.}
\end{eqnarray}%
Finally, by introducing a unit vector $\mathbf{\tilde{n}}$ defined by 
\[
\mathbf{\tilde{n}}=\widetilde{z}^{\dagger }\mathbf{\sigma }\widetilde{z} 
\]%
and by using

\[
\frac{1}{4}\left\vert \mathbf{\nabla \tilde{n}}\right\vert ^{2}=|\mathbf{%
\nabla }\tilde{z}|^{2}-\left\vert \mathbf{v}^{s}\right\vert ^{2} 
\]%
the low-energy effective Lagrangian reduces to 
\begin{equation}
\mathcal{L}_{\mathrm{eff}}=\frac{1}{8\tilde{g}}(\mathbf{\nabla \tilde{n})}%
^{2}+\frac{\tilde{g}}{2\pi ^{2}}(\mathbf{E}^{s})^{2}+iA_{0}^{s}\mathcal{K}%
_{0}^{s}+\mathcal{L}_{h}  \label{AFMCoulomb}
\end{equation}%
where 
\begin{eqnarray*}
\mathcal{K}_{0}^{s} &\equiv &\frac{1}{\pi }\epsilon _{0\nu \lambda }\mathbf{%
\partial }_{\nu }v_{\lambda } \\
&=&\frac{1}{4\pi }\epsilon _{0\nu \lambda }\mathbf{\tilde{n}\cdot \partial }%
^{\nu }\mathbf{\tilde{n}\times \partial }^{\lambda }\mathbf{\tilde{n}.}
\end{eqnarray*}

This low-energy Lagrangian describes how the bosonic holons, via $\mathcal{L}%
_{h}$, and spin twists, with topological charge density $\mathcal{K}_{0}^{s}$%
, are coupled to a Maxwell gauge field $A^{s}$ \emph{with the
\textquotedblleft photon velocity\textquotedblright\ $c=\infty $}, that is,
in the absence of $\left\vert \mathbf{B}^{s}\right\vert ^{2}$. The only
effect of such a non-relativistic gauge field is then to induce a 2D Coulomb
interaction between two types of charged particles, including holons and
spin twists characterized by $\mathcal{K}_{0}^{s}$. Noting $\mathbf{E}%
^{s}=-\nabla A_{0}^{s}$ and integrating out $A_{0}^{s}$ in (\ref{AFMCoulomb})%
$,$ a potential term will emerge in the effective action as 
\begin{equation}
V=q_{h}^{2}\int d^{2}\mathbf{r}d^{2}\mathbf{r}^{\prime }\ln \left\vert 
\mathbf{r}-\mathbf{r}^{\prime }\right\vert \left( \rho _{h}+\mathcal{K}%
_{0}^{s}\right) (\mathbf{r})\left( \rho _{h}+\mathcal{K}_{0}^{s}\right) (%
\mathbf{r}^{\prime })  \label{V}
\end{equation}%
in which $\rho _{h}=h^{\dagger }h$ and $q_{h}^{2}=\pi ^{3}/\tilde{g}^{2},$
accompanied by a \emph{charge neutral} constraint enforced in the
thermodynamic limit on the low energy states of such a 2D Coulomb gas
system, namely

\begin{equation}
\int d^{2}\mathbf{r}\left[ \rho _{h}(\mathbf{r})+\mathcal{K}_{0}^{s}(\mathbf{%
r})\right] =0.  \label{dualflux}
\end{equation}%
Thus, a holon has to be \textquotedblleft confined\textquotedblright\ to a
spin twist, satisfying $1+\int d^{2}\mathbf{r}\frac{1}{2\pi }\mathbf{\tilde{n%
}}\cdot \partial _{x}\mathbf{\tilde{n}}\times \partial _{y}\mathbf{\tilde{n}}%
=0,$ which leads to the quantization condition of the winding number of the
unit vector $\left\{ \mathbf{\tilde{n}(r})\right\} $ in spin space as
follows 
\begin{eqnarray}
\mathcal{Q}^{s} &\equiv &\int d^{2}\mathbf{r}\frac{1}{4\pi }\mathbf{\tilde{n}%
}\cdot \partial _{x}\mathbf{\tilde{n}}\times \partial _{y}\mathbf{\tilde{n}}
\nonumber \\
&=&-\frac{1}{2}.  \label{Q}
\end{eqnarray}%
Namely, each holon will be bound to a \textquotedblleft
meron\textquotedblright , which is a spin twist of the unit vector $\mathbf{%
\tilde{n}}$ whose winding number is half of that for a Skyrmion.

According to the condition (\ref{dualflux}), one expects to find equal
number of holons and (anti)merons at low temperatures, which are paired by
the logarithmic-attractive interaction in (\ref{V}). An unpaired holon or
(anti)meron will cost a logarithmically divergent energy and thus is
forbidden to appear. In other words, in the AF phase, a bare holon can not
exist alone, but has to be always confined to a spin topological
configuration (meron). Such an effect in the spin ordered phase is called
the \textquotedblleft \emph{holon confinement}\textquotedblright . Note that
a holon itself will also carry a spin vortex according to (\ref{operators}),
the composite object formed by the holon-meron pair actually corresponds to
a spin dipolar configuration in the \emph{real} spin space, as previous
identified in the phase string model \cite{kou,kou1}. Since the (anti)meron
is a semiclassical object without a coherent quantum dynamics, the dipole as
a bound pair of a holon and a (anti)meron normally cannot move coherently
either. That is, the holon will be \emph{self-trapped} near the core of the
meron in space and the translation symmetry is spontaneously broken.

With the increase of doping, \emph{i.e.,} the number of holon-antimeron
dipoles, one expects to see a screening effect on the confining potential $V$%
. It has been previously found that eventually a confinement-deconfinement
transition can take place beyond some critical doping concentration, where
the screened 2D Coulomb interaction becomes short-ranged \cite{kou,kou1}.
Once the bosonic holons are free, they will experience a Bose condensation
and the resulting phase is an SC state as to be discussed in the following
section. In the SC phase, there exists a duality correspondence of the
quantization condition (\ref{Q}), which will ensures the flux quantization
condition there. Correspondingly, (\ref{Q}) may be called a \emph{dual
flux-quantization condition}.

Finally we remark that the hole self-trapping at low doping, discussed in
the present work, is in contrast to a conventional picture for single hole
moving in the AF background based on the numerical studies of the $t-J$
model \cite{tklee}. In the latter case, the doped hole is found to have
finite spectral weight and a coherent dispersion with the bandwidth
comparable to $J.$ The discrepancy may arise from the small sample sizes in
exact diagonalization calculations: The phase string effect, which leads to
the mutual Chern-Simons gauge fields, starts to play the role of
self-localization only when the sample sizes become larger than the
localization length scales \cite{1-hole}. Further investigations from both
analytic and numerical approaches to clarify this issue are needed. Possible
experimental implications of self-trapping for lightly doped cuprate have
been previously discussed in the phase string model \cite{kou1}.

\subsection{Meissner effect and spinon confinement in SC phase}

Now let us consider the other ordered phase with the Bose condensation of
holons, $\left\langle h\right\rangle \neq 0$, whose ground state is a
superconducting one \cite{weng-LG} with the Meissner effect and charge $2e$
minimal flux quantization as shown below.

With $\left\langle h\right\rangle \neq 0$, $\mathcal{L}_{h}$ in (\ref%
{continous}) reduces to%
\begin{equation}
\mathcal{L}_{h}=i\rho _{h}(\partial _{0}\phi _{h}-A_{0}^{s})+\frac{\rho _{h}%
}{2m_{h}}\left( \mathbf{\nabla }\phi _{h}-\mathbf{A}^{s}-\mathbf{A}%
^{e}\right) ^{2}  \label{lh-sc}
\end{equation}%
\newline
by writing $h(\mathbf{r})=\sqrt{\rho _{h}}e^{i\phi _{h}(\mathbf{r})}$. The
Chern-Simons term (\ref{continous}) can be rewritten as%
\begin{equation}
\mathcal{L}_{\mathrm{CS}}=-\frac{i}{\pi }\mathbf{A}^{s}\cdot (\mathbf{E}%
^{h}\times \hat{\mathbf{z}})+\frac{i}{\pi }A_{0}^{s}\mathbf{B}^{h}\cdot \hat{%
\mathbf{z}}  \label{cs-cs}
\end{equation}%
by introducing the \textquotedblleft electric\textquotedblright\ field $%
\mathbf{E}^{h}=\partial _{0}\mathbf{A}^{h}-\nabla A_{0}^{h}$ and
\textquotedblleft magnetic\textquotedblright\ field $\mathbf{B}^{h}=$ $%
\nabla \times \mathbf{A}^{h}$ for the vector potential $\mathbf{A}^{h}$.

Firstly, the \textquotedblleft magnetic\textquotedblright\ field $B^{h}=%
\mathbf{B}^{h}\cdot \hat{\mathbf{z}}$ can be determined after integrating
out $A_{0}^{s}$ in the partition function and one obtains the condition 
\begin{equation}
B^{h}=\mathbf{B}^{h}\cdot \hat{\mathbf{z}}=\pi \rho _{h}  \label{bh}
\end{equation}%
which is uniform and fixes the spatial component $\mathbf{A}^{h}$, such that 
$\mathbf{E}^{h}=-\nabla A_{0}^{h}$. Secondly, after integrating out $\mathbf{%
A}^{s},$ the resulting effective Lagrangian takes the following form 
\begin{equation}
\mathcal{L}_{\mathrm{eff}}=\mathcal{L}_{s}+\left( \frac{m_{h}}{2\pi ^{2}\rho
_{h}}\right) \left\vert \mathbf{E}^{h}\right\vert ^{2}-iA_{0}^{h}\mathcal{Q}%
^{h}  \label{eff-sc}
\end{equation}%
in which 
\[
\mathcal{Q}^{h}\equiv \frac{1}{\pi }\epsilon ^{0\nu \lambda }\partial _{\nu
}\left( \partial _{\lambda }\phi _{h}-A_{\lambda }^{e}\right) . 
\]

Finally, we integrate out $A_{0}^{h}$ in (\ref{eff-sc}). For our purpose,
instead of using the continuous version (\ref{continous}) of $\mathcal{L}%
_{s} $, we shall use a simpler but more precise form of the term involving $%
A_{0}^{h}$ based on the original $L_{s}$ defined in (\ref{Lagrangian-CS}),
which reads

\[
\mathcal{L}_{s}=-iA_{0}^{h}\rho _{s}(\mathbf{r})+\mathcal{L}_{s}\left(
A_{0}^{h}=0\right) 
\]%
in which $\rho _{s}(\mathbf{r})=\rho _{\uparrow }(\mathbf{r})-\rho
_{\downarrow }(\mathbf{r})$ with $\rho _{\uparrow }(\mathbf{r})$[$\rho
_{\downarrow }(\mathbf{r})$] denotes the density of up (down) spinons. Then,
after integrating out $A_{0}^{h},$ one obtains the following effective
action in (2+1)-dimensional Euclidean space

\[
S_{\mathrm{eff}}=\int d^{3}x_{\mu }\left[ \mathcal{L}_{s}\left(
A_{0}^{h}=0\right) 
\right] +\int dx_{0}V_{\mathrm{SC}} 
\]%
where 
\begin{equation}
V_{\mathrm{SC}}=q_{s}^{2}\int d^{2}\mathbf{r}d^{2}\mathbf{r}^{\prime }\ln
\left\vert \mathbf{r}-\mathbf{r}^{\prime }\right\vert \left( \rho _{s}+%
\mathcal{Q}^{h}\right) (\mathbf{r})\left( \rho _{s}+\mathcal{Q}^{h}\right) (%
\mathbf{r}^{\prime })  \label{vsc}
\end{equation}%
with $q_{s}^{2}=\frac{\pi \rho _{h}}{4m_{h}}$. Similar to the case in the AF
phase, in the thermodynamic limit, there is a charge neutral condition
enforced as follows: 
\begin{eqnarray}
0 &=&\int d^{2}\mathbf{r}\left[ \rho _{s}(\mathbf{r})+\mathcal{Q}^{h}(%
\mathbf{r})\right]  \nonumber \\
&=&N_{\uparrow }-N_{\downarrow }+\left( 2N_{\mathrm{vor}}-\frac{\Phi ^{e}}{%
\pi }\right) ,  \label{cn-SC}
\end{eqnarray}%
in which $N_{\uparrow }$ ($N_{\downarrow })$ is the total number of spin-up
(spin-down) spins, $N_{\mathrm{vor}}=\frac{1}{2\pi }\int d^{2}\mathbf{r}%
\epsilon ^{\alpha \beta }\partial _{\alpha }\partial _{\beta }\phi _{h}$
denotes the total number of $2\pi $ vortices in the holon field, and $\Phi
^{e}$ is the total external magnetic flux $\Phi ^{e}=\int d^{2}\mathbf{r}%
\epsilon ^{0\nu \lambda }\partial _{\nu }A_{\lambda }^{e}$.

Since $N_{\uparrow }$, $N_{\downarrow },$ and $N_{\mathrm{vor}}$ are all
quantized to be integers, we find the minimal \emph{flux quantization
condition} 
\begin{equation}
\left\vert \Phi _{\mathrm{\min }}^{e}\right\vert =\frac{\Phi _{0}}{2}
\label{fq}
\end{equation}%
where $\Phi _{0}=2\pi $ ($=hc/e$ in full units) is the flux quantum for a
charge $e$ system.

Therefore, the external magnetic flux is not allowed to present in the bulk
(i.e., the Meissner effect) unless it is quantized in multiples of half flux
quanta given in (\ref{fq}). In particular, for a magnetic flux quantized at
the minimal half flux quantum $\Phi _{\mathrm{\min }}^{e}$, there must be a
spinon trapped near the vortex core according to (\ref{cn-SC}) by noting
that the \textquotedblleft charge\textquotedblright\ $2N_{\mathrm{vor}}$ of
the vortices produced by holon field is always in units of $2\pi $ (i.e., $%
\Phi _{0}$).

However, free spinons are not allowed in the bulk in the absence of the
external magnetic flux. Indeed, for $\Phi ^{e}=0,$ the \textquotedblleft
charge neutral\textquotedblright\ condition (\ref{cn-SC}) reduces to $%
N_{\uparrow }-N_{\downarrow }+2N_{\mathrm{vor}}=0$. As the result, a single
spinon excitation, with $S^{z}=\left( N_{\uparrow }-N_{\downarrow }\right)
/2=\pm 1/2,$ will violate the \textquotedblleft charge
neutral\textquotedblright\ condition, which in fact will cost a
logarithmically divergent energy as each spinon behaves like a half vortex.
Hence, in the superconducting state, the spinons-vortices must be always
paired up (confined) in the bulk by the logarithmic force given in (\ref{vsc}%
). To be noted, not only the spinons with different spin indices (up/down)
can pair up, those with the same spin indices (up/up and down/down) can also
pair up to satisfy the charge neutral condition by involving a holon phase
vortex with $N_{\mathrm{vor}}\neq 0$, \emph{e.g.,} $N_{\mathrm{vor}%
}=1,N_{\uparrow }=2,N_{\downarrow }=0$ or $N_{\mathrm{vor}}=-1,N_{\uparrow
}=0,N_{\downarrow }=2$. Since $N_{\mathrm{vor}}$ does not appear in the rest
of the action, these two excitations of $S^{z}=\pm 1$ are energy degenerate
with the state $S^{z}=0,$ $N_{\mathrm{vor}}=0$, to form $\mathbf{S}=1$
triplet spin excitations, consistent with the spin rotation symmetry
generally demonstrated before. For the same reason, the single spinon bound
to a magnetic vortex quantized at $\frac{\Phi _{0}}{2}$ should have a free
moment with $S^{z}=\pm 1/2$, because of the freedom introduced by $N_{%
\mathrm{vor}}$.

Thus, the superconducting phase in the present mutual-Chern-Simons theory is
characterized by the holon condensation and spinon (logarithmic)
confinement. We have seen that the fractionalization of spins (a single
spinon) does not directly appear in the bulk low-lying excitation spectrum,
but does show up in a magnetic vortex core \cite{weng-LG}. The deconfinement
of spinon-vortex pairs will eventually occur at the superconducting
transition temperature $T_{c}$ \cite{weng-LG,ming}. Finally, we point out
that the symmetry of the superconducting order parameter, which is expressed
in terms of the electron operator (\ref{ps}), is d-wave like as discussed
previously in Ref. \cite{yi}.

\subsection{Mutual duality of two phases}

The doping effect and the interplay between charge and spin degrees of
freedom are characterized by a mutual-Chern-Simons gauge structure in this
model, as discussed in previous sections. The mutually dual characteristics
of these two phases are summarized by the following table.

\begin{center}
$%
\begin{array}{|c|c|c|}
\hline
& \text{AF} & \text{SC} \\ \hline
\text{Bose condensation} & <\mathbf{z}>\neq 0 & <h>\neq 0 \\ \hline
\text{Coulomb gauge field} & A_{0}^{s} & A_{0}^{h} \\ \hline
\begin{array}{c}
``\text{charged\textquotedblright\ particle of} \\ 
\text{Coulomb gauge field}%
\end{array}
& \text{holon} & \text{spinon} \\ \hline
\begin{array}{c}
\text{external source of } \\ 
\text{Coulomb gauge field}%
\end{array}
& \text{meron} & \text{magnetic flux} \\ \hline
``\text{charge neutral\textquotedblright\ object} & \text{holon-meron pair}
& 
\begin{array}{c}
\text{a. \ \ spinon pair} \\ 
\text{b.\ magnetic flux}+\text{ a spinon }%
\end{array}
\\ \hline
\text{dual flux quantization} & |\mathcal{Q}^{s}|\text{ }=\frac{1}{2} & 
\left\vert \Phi _{\mathrm{\min }}^{e}\right\vert =\frac{\Phi _{0}}{2}=\frac{%
hc}{2e} \\ \hline
\text{dual Meissner effect} & \text{holon confinement} & 
\begin{array}{c}
\text{a. spinon confinement} \\ 
\text{b. spinon bound to magnetic flux}%
\end{array}
\\ \hline
\end{array}%
$
\end{center}

We have shown that, at low doping, the spinon condensation leads to a spin
AF order and forces a \textquotedblleft confinement\textquotedblright\ on
the holon part, making holons self-localized to ensure the AFLRO. On the
other hand, at a higher doping, the condensation of bosonic holons forces a
\textquotedblleft confinement\textquotedblright\ on the spinon part,
resulting an SC phase coherence. \ 

There are several distinctions between the two ordered phases. In the AF
phase, the spinons condensate is a kind of two-component \textquotedblleft
superfluid\textquotedblright . Consequently the global symmetry is broken
from \textrm{SU(2)} to \textrm{U(1).} In contrast, the ground state in the
SC phase is a condensation of a scalar field - holons. As a result the
global \textrm{U(1)} symmetry of the charge part is broken.

Besides such a fundamental distinction, two phases share some common
features originated from the duality in the mutual-Chern-Simons gauge
structure. In both the AF and SC phases, there exist induced Maxwell terms
that have only \textquotedblleft electric field strengths\textquotedblright\
without the Lorentz invariance. There are \textquotedblleft
charges\textquotedblright\ coupling to these Coulomb gauge fields, including
quantum particles (holons and spinons) and external sources without quantum
dynamics. The \textquotedblleft charge neutral\textquotedblright\ condition
and the 2D Coulomb interaction among the \textquotedblleft
charged\textquotedblright\ objects lead to dual Meissner effects: In the SC
case, an external magnetic flux as an external source must be quantized, and
in order to realize a minimal quantum $hc/2e,$ a single spinon must be bound
to such a magnetic flux to form a \textquotedblleft charge
neutral\textquotedblright\ object; In the AFLRO state, as an external
source, a (anti)meron is allowed as a topological excitation from the spinon
condensate with a quantized winding number $\left|\mathcal{Q}^{s}\right|=%
\frac{1}{2}$ and a holon cannot live alone and must be bound to such an
external source to form a \textquotedblleft charge
neutral\textquotedblright\ object with a spin dipolar configuration.

Finally we emphasize that the mutual-Chern-Simons theory in this work
involves a mutual duality between the charge and spin degrees of freedom
rather than a usual duality. A usual dual description has been also widely
used \cite{Balents,dhlee,fra,za} in studying the doped Mott insulators,
which deals with an ordered phase and the transition to a disordered phase
in terms of the corresponding topological defects on the dual lattice. In a
conventional dual-theory description, normally the AF and SC phases are not
directly related. By contrast, in the mutual duality discussed in the
present theory, the vortices of one species (holon/spinon) under
condensation are themselves quantum objects of another species
(spinon/holon), and \emph{two} ordered phases, i.e., AF and SC states, can
be naturally unified together \cite{kou}. A similar duality at low doping
has been also investigated \cite{ng1} by starting from the slave-fermion
approach \cite{ng}. Based on the same phase string decomposition (\ref{ps})
but with a slightly different mean-field decoupling (see Ref. \cite{remark1}%
), a mutual duality between the AF and SC states has been also discussed
recently in Ref. \cite{wang}.

\section{conclusion}

In this paper, we studied a new class of nontrivial (2+1)-dimensional gauge
field structure - the mutual-Chern-Simons theory. The Lagrangian of such a
mutual-Chern-Simons theory is derived as an effective low-energy description
of the phase-string model for doped Mott insulators. This effective
Lagrangian retains the full symmetries of parity, time-reversal, and global
SU(2) spin rotation, in contrast to the conventional Chern-Simons theories
where first two symmetries are usually broken.

The mutual-Chern-Simons theory as a minimal model for doped Mott insulators
has a unique mutual duality structure. Two ordered phases found in this
theory, the AF and SC states, are connected by dual Meissner/flux
quantization effects and dual confinement/deconfinement. Namely, holons
become vortices in the spinon condensed AF phase and spinons become vortices
in the holon condensed SC state. The former leads to the holon confinement
(a holon bound to a spin meron twist to form a \textquotedblleft
neutral\textquotedblright\ dipolar structure) and the latter leads to the
spinon confinement and flux quantization (a spinon bound to a magnetic flux
quantized at $hc/2e$ to form a \textquotedblleft neutral\textquotedblright\
object in the holon condensate).

Such a mutual duality structure between the charge and spin degrees of
freedom determines the essential competition between two degrees of freedom
and provides driving forces for phase transitions to each other or to other
disordered phases. Dual confinement means that there is no true spin-charge
separation is present in these ordered phases since one species
(spinon/holon) is always confined at low temperatures while the other
(spinon/holon) is condensed. But the dual deconfinement will play an
essential role in the transitions to disordered phases or at the boundary
between two ordered phases where a quantum critical point may exist \cite%
{kou,QCP}. The systematic evolution of the phase diagram at low doping is
currently under investigation based on the mutual-Chern-Simons Lagrangian.

In the future we will also consider some additional relevant terms which
have not been taken into consideration in the present \emph{minimal} model.
As previously shown \cite{yi}, there generally exists a residual attractive
interaction between holons and spinons within the $t-J$ model, which should
be included when one considers the nodal (d-wave) fermionic quasiparticle
excitations as \textquotedblleft collective\textquotedblright\ modes in the
SC phase. In principle, besides spinon confinement, there also exists a
holon-spinon confinement in the SC phase of the phase string model, since
single spinon or holon excitation is not allowed \cite{yi}. How the
fermionic nodal quasiparticles can be naturally described in the mutual
Chern-Simons framework will be a central issue to address in a next study,
where the attractive interaction between holons and spinons beyond the phase
string model should be properly incorporated in order to get a correct
excitation spectrum \cite{yi}. We do not expect a qualitative modification
on the present results of the minimal model by including such a term, since
the quasiparticles as bound states of holons and spinons are independent, to
leading order approximation, of those spinon excitations which are confined
to form integer neutral spin excitaions discussed in the present work.

\begin{acknowledgments}
We thank helpful discussions with L. Balents, M. L. Ge, V. N. Muthukumar, T.
K. Ng, D. N. Sheng, X. G. Wen, Y. S. Wu, J. Zaanen, and Y. Zhou. The authors
also acknowledge partial support from the grants of NSFC, the grant no.
104008 and SRFDP from MOE. S.P.K. acknowledges support from Beijing Normal
University.
\end{acknowledgments}

\appendix

\section{Single-valueness of $\Phi ^{h}(\mathbf{r},\protect\tau )$}

In the transformation (\ref{su2}), $\Phi ^{h}(\mathbf{r},\tau )$ is required
to be single-valued with mod $2\pi $ in order to ensure the single-valueness
of the spin operators$.$ Then according to (\ref{su2-1}), it imposes a
constraint on $A_{\mu }^{h}$, i.e.,

\begin{equation}
\Delta \Phi ^{h}|_{C}=2\oint_{C}A_{\mu }^{h}dx_{\mu }=2n\pi ,\text{ }n\in 
\mathbb{Z}  \label{ConstraintA}
\end{equation}%
in which $dx_{\mu }$ is the tangential differential vector of an arbitrary
loop $C$ in (2+1) dimensions.

Since the gauge field $A_{\mu }^{h}$ is an independent dynamic variable, the
constraint (\ref{ConstraintA}) would be generally \emph{violated}. However,
we shall prove below that all the $A_{\mu }^{h}$ configurations that violate
(\ref{ConstraintA}) have vanishing contribution to the partition function,
which is consistent to the topological constraint (\ref{constraint2}) in the
Hamiltonian formalism.

\begin{figure}[tbp]
\begin{center}
\includegraphics[width=2.5in] {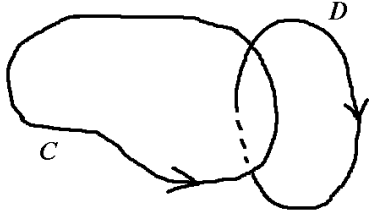}
\end{center}
\caption{A vortex ring phase $\protect\theta $ is defined such that $\protect%
\oint_{D}d\protect\theta =2\protect\pi M$ for any circuit $D$ winding around
the loop $C$ once. }
\label{vortexring}
\end{figure}

First of all, for an arbitrary loop $C$ in (\ref{ConstraintA}), we may
introduce a vortex ring phase configuration $e^{i\theta (x_{\mu })}$ with an
arbitrary winding number $M$ ($M\in \mathbb{Z}$), which satisfies $%
\oint_{D}d\theta =2\pi M$ for any circuit $D$ that winds around $C$ once, as
shown in Fig. \ref{vortexring}. This singularity in $e^{i\theta (x,t)}$ can
be clearly expressed by 
\begin{equation}
\epsilon ^{\mu \nu \lambda }\partial _{\nu }\partial _{\lambda }\theta
(x)=2\pi M\oint_{C}dy_{\mu }(C)\delta ^{(3)}\left( x_{\mu }-y_{\mu
}(C)\right) ,\text{ }M\in \mathbb{Z}  \label{vortex}
\end{equation}%
in which $y_{\mu }(C)$ represents the coordinates on the loop $C$.

Then, we can make a singular gauge transformation in terms of such a phase $%
\theta (x_{\mu })$ as 
\begin{equation}
\widetilde{h}=e^{i\theta }h,\text{ }\widetilde{A}_{\mu }^{s}=A_{\mu
}^{s}+\partial _{\mu }\theta .
\end{equation}%
Lagrangian $\mathcal{L}_{h}$ and $\mathcal{L}_{s}$ remain invariant, but the
mutual-Chern-Simons term in (\ref{continous}) changes as 
\begin{equation}
\widetilde{\mathcal{L}}_{CS}=\mathcal{L}_{CS}+\frac{i}{\pi }\partial _{\mu
}\theta \epsilon ^{\mu \nu \lambda }\partial _{\nu }A_{\lambda }^{h}
\end{equation}%
such that the total action is transformed as%
\begin{eqnarray}
\widetilde{S}_{\mathrm{eff}} &=&S_{\mathrm{eff}}+\int d^{3}x_{\mu }\frac{i}{%
\pi }\partial _{\mu }\theta \epsilon ^{\mu \nu \lambda }\partial _{\nu
}A_{\lambda }^{h}  \nonumber \\
&=&S_{\mathrm{eff}}+\int d^{3}x_{\mu }\frac{i}{\pi }A_{\mu }^{h}\epsilon
^{\mu \nu \lambda }\partial _{\nu }\partial _{\lambda }\theta  \nonumber \\
&=&S_{\mathrm{eff}}+i2M\oint_{C}A_{\mu }^{h}dx_{\mu }.
\end{eqnarray}

Therefore, the partition function can be written as 
\begin{eqnarray}
Z &=&\int D[..]\exp \left( -S_{\mathrm{eff}}-2iM\oint_{C}A_{\mu }^{h}dx_{\mu
}\right)  \nonumber \\
&=&Const\cdot \sum_{M\in \mathbb{Z}}\int D[..]\exp \left( -S_{\mathrm{eff}%
}-iM\oint_{C}2A_{\mu }^{h}dx_{\mu }\right)
\end{eqnarray}%
where $D[..]$ stands for the functional integrations over all the fields, $h$%
, $h^{\ast }$, $z_{\sigma }$, $z_{\sigma }^{\ast }$, $A_{\mu }^{h}$, and $%
A_{\mu }^{s}$. The summation over $M$ directly lead to the constraint (\ref%
{ConstraintA}) for an arbitrary loop $C$.

\begin{thebibliography}{99}
\bibitem{anderson} P. W. Anderson, Science \textbf{235}, 1196 (1987).

\bibitem{anderson1} Z. Zou and P. W. Anderson, Phys. Rev. B \textbf{37,} 627
(1988).

\bibitem{krs} S.A. Kivelson, D.S. Rokhsar, and J.R. Sethna, Phys. Rev. B 
\textbf{35}, 8865 (1987).

\bibitem{an1} P.W. Anderson, \emph{The Theory of Superconductivity in the
High-T}$_{c}$\emph{\ Cuprates} (Princeton University Press, 1997).

\bibitem{fra1} \emph{see, }E. Fradkin, \emph{Field Theories of Condensed
Matter Systems} (Addison--Wesley, New York, 1991).

\bibitem{u1} L. B. Ioffe and A. I. Larkin, Phys. Rev. B\textbf{39}, 8988
(1989).

\bibitem{lee} N. Nagaosa and P. A. Lee, Phys. Rev. Lett. \textbf{64}, 2450
(1990); P. A. Lee and N. Nagaosa, Phys. Rev. B\textbf{46}, 5621 (1992).

\bibitem{su2} Xiao-Gang Wen and P.A. Lee, Phys. Rev. Lett. \textbf{76}, 503
(1996); Patrick A. Lee, Naoto Nagaosa, Tai-Kai Ng, Xiao-Gang Wen, Phys. Rev.
B\textbf{57}, 6003 (1998).

\bibitem{z2} T. Senthil and Matthew P.A. Fisher, Phys. Rev. B\textbf{62},
7850 (2000); Phys. Rev B\textbf{63}, 134521 (2001).

\bibitem{wie} P. B. Wiegmann, Phys. Rev. Lett., \textbf{60}, 821 (1988).

\bibitem{shankar} R. Shankar, Phys. Rev. Lett. \textbf{63}, 203 (1989); R.
Shankar, Nucl. Phy.\textbf{\ }B\textbf{330}, 433 (1990).

\bibitem{lee2} P.A. Lee, Phys. Rev. Lett\textit{.} \textbf{63, }680 (1989).

\bibitem{weng1} Z. Y. Weng, Phys. Rev. Lett. \textbf{66}, 2156 (1991).

\bibitem{su} P. A. Marchetti, Z. B. Su and L. Yu, Phys. Rev. B\textbf{58},
5808 (1998); Mod. Phys. Lett. B\textbf{12}, 173 (1998).

\bibitem{ng} T. K. Ng, Phys. Rev. B\textbf{52}, 9491 (1995); Phys. Rev.
Lett. \textbf{82}, 3504 (1999).

\bibitem{aa} D.P. Arovas and A. Auerbach, Phys. Rev. B 38, 316 (1988); A.
Auerbach, \emph{Interacting Electrons and Quantum Magnetism,}
(Sprinber-Verlag, 1994).

\bibitem{wiegmann} P.\ B. Wiegmann, Phys. Rev. Lett. \textbf{65}, 2070
(1990).

\bibitem{rod} J. P. Rodriguez and B. Doucot, Phys. Rev. B\textbf{42}, 8724
(1990).

\bibitem{tik} A.M. Tikofsky, R.\ B. Laughlin, and Z. Zou, Phys. Rev. Lett., 
\textbf{69}, 3670 (1992).

\bibitem{weng-semion} Z. Y. Weng, D. N. Sheng, and C. S. Ting, Phys. Rev. B%
\textbf{49}, 607 (1994).

\bibitem{phase-string1} Z. Y. Weng, D. N. Sheng, and C. S. Ting, Phys. Rev.
Lett. \textbf{80}, 5401 (1998); Phys. Rev. B\textbf{59}, 8943 (1999).

\bibitem{ps00} Z. Y. Weng, D. N. Sheng, and C. S. Ting, Phys. Rev. B\textbf{%
52}, 637 (1995); Mod. Phys. Lett. B\textbf{8}, 1353 (1994).

\bibitem{phase-string0} Z. Y. Weng, D. N. Sheng, Y. C. Chen, and C. S. Ting,
Phys. Rev. B \textbf{55}, 3894 (1997); D. N. Sheng, Y. C. Chen, and Z. Y.
Weng, Phys. Rev. Lett. \textbf{77}, 5102 (1996).

\bibitem{wang} Qiang-Hua Wang, Phys. Rev. Lett. \textbf{92}, 057003 (2004);
Chin. Phys. Lett. \textbf{20}, 1582 (2003).

\bibitem{remark1} The usual U(1) symmetry similar to the slave-boson
approach is broken in the bosonic RVB mean-field state upon doping, by the
hopping term \cite{phase-string1}, in the phase string formalism. On the
other hand, without considering the RVB mean-field saddle point, such a U(1)
gauge field has been integrated out in Ref. \cite{wang}, leading to a
slightly different gauge field description based on the same phase string
decomposition.

\bibitem{nayak2} C. Nayak, Phys. Rev. Lett. \textbf{86}, 943 (2001); M.
Oshikawa, Phys. Rev. Lett. \textbf{91}, 199701 (2003).

\bibitem{weng-LG} V.N. Muthukumar, Z.Y. Weng, Phys. Rev. B\textbf{65},
174511 (2002).

\bibitem{kou} S. P. Kou and Z. Y. Weng, Phy. Rev. Lett. \textbf{90}, 157003
(2003).

\bibitem{kou1} S. P. Kou and Z. Y. Weng, cond-mat/0402327.

\bibitem{confine} A.M. Polyakov, Nucl. Phys. \textbf{B 120}, 429 (1977).
A.M. Polyakov, \emph{Gauge fields and strings} (Harwood Academic Publishers,
London, 1987).

\bibitem{nayak1} C. Nayak, Phys. Rev. Lett., \textbf{85}, 178 (2000).

\bibitem{laughlin} R. B. Laughlin, Science \textbf{242} (1988) 525; Phys.
Rev. Lett. \textbf{60, }2677 (1988).

\bibitem{wwz} X. G. Wen, F. Wilczek, A. Zee, Phys. Rev. B39, 11413 (1989);
X.G. Wen and A. Zee, Nucl. Phys\textit{.} B\textbf{326, }619 (1989).

\bibitem{wilczek} see, F. Wilczek, \emph{Fractional Statistics and Anyon
Superconductivity}, (World Scientific, 1990); and the references therein.

\bibitem{frac} F. Wilczek, Phys. Rev. Lett. \textbf{69} 132 (1992).

\bibitem{Balents} L. Balents, M. P. A. Fisher, C. Nayak, Phys. Rev. B\textbf{%
60} 1654 (1999); Phys. Rev. B \textbf{61} 6307 (2000).

\bibitem{dhlee} D. H. Lee, Phys. Rev. Lett. \textbf{88}, 227003 (2002).

\bibitem{fra} M. Franz, Z. Tesanovic, O. Vafek, Phys. Rev. B\textbf{66},
054535 (2002); I. F. Herbut and D. J. Lee, Phys. Rev. B\textbf{68, } 104518
(2003).

\bibitem{za} J. Zaanen, Z. Nussinov, S.I. Mukhin, Annals of Physics 310
(2004) 181.

\bibitem{ming} M. Shaw, Z. Y. Weng, and C. S. Ting, Phys. Rev. B\textbf{68},
014511 (2003).

\bibitem{read} N. Read and S. Sachdev, Phys. Rev. Lett. 62, 1694, (1989).

\bibitem{cha} S. Chakravarty, B. I. Halperin and D. R. Nelson, Phy. Rev.
Lett. 60, 1057 (1988); Phy. Rev. B 39, 2344 (1989).

\bibitem{tri} A. M. Tsvelik, \emph{Quantum Field Theory in Condensed Matter
Physics, }(Cambridge University Press, 1995).

\bibitem{sach} S. Sachdev, \emph{Quantum Phase Transitions}, (Cambridge
University Press, 1999).

\bibitem{tklee} \emph{see,} T.K. Lee, C.M. Ho, and, N. Nagaosa, Phys. Rev.
Lett. \textbf{90}, 067001 (2003), and the references therein.

\bibitem{1-hole} Z. Y. Weng, D. N. Sheng, and C. S. Ting, Phys. Rev. B%
\textbf{63}, 075102 (2001).

\bibitem{yi} Y. Zhou, V. N. Muthukumar, and Z. Y. Weng, Phys. Rev. B \textbf{%
67}, 064512 (2003); Z. Y. Weng, D. N. Sheng, and C.S. Ting, Phys. Rev. B 
\textbf{61}, 12328 (2000).

\bibitem{ng1} T. K. Ng, Int. J. Mod. Phys. B\textbf{14}, 349 (2000).

\bibitem{QCP} T. Senthil, Ashvin Vishwanath, Leon Balents, Subir Sachdev, M.
P. A. Fisher, Science \textbf{303}, 1490 (2004).
\end{thebibliography}
\end{document}